# Rapid Anxiety Reduction (RAR):
# A unified theory of humor

Adam Safron
Indiana University

**Abstract**
Here I propose a novel theory in which humor is the feeling of Rapid Anxiety Reduction (RAR). According to RAR, humor can be expressed in a simple formula: *-d(A)/dt*. RAR has strong correspondences with False Alarm Theory, Benign Violation Theory, and Cognitive Debugging Theory, all of which represent either special cases or partial descriptions at alternative levels of analysis. Some evidence for RAR includes physiological similarities between hyperventilation and laughter and the fact that smiles often indicate negative affect in non-human primates (e.g. fear grimaces where teeth are exposed as a kind of inhibited threat display). In accordance with Benign Violation Theory, if humor reliably indicates both a) anxiety induction, b) anxiety reduction, and c) the time-course over which anxiety is reduced, then the intersection of these conditions productively constrains inference spaces over latent mental states with respect to the values and capacities of the persons experiencing humor. In this way, humor is a powerful cypher for understanding persons in both individual and social contexts, with far-reaching implications. Finally, if humor can be expressed in such a simple formula with clear ties to phenomenology, and yet this discovery regarding such an essential part of the human experience has remained undiscovered for this long, then this is an extremely surprising state of affairs worthy of further investigation. Towards this end, I propose an analogy can be found with consciousness studies, where in addition to the "Hard problem" of trying to explain humor, we would do well to consider a "Meta-Problem" of why humor seems so difficult to explain, and why relatively simple explanations may have eluded us for this long. *(Please note: RAR was conceived in 2008, and last majorly updated in 2012.)*

## Table of Contents

## Rapid Anxiety Reduction (RAR)

RAR can be stated simply as follows: ***Humor is the feeling of rapid anxiety reduction.***

Considered as a continuous function, the moment-to-moment experience of humor can be quantified as the negative first (or possibly second[4]) derivative of anxiety with respect to time:

***Humor = -d(A)/dt***

Less simply, both reduction of anxiety as well as rapidity of change are necessary and sufficient conditions for producing the subjective experience of humor. According to RAR, humor is experienced as pleasurable and reinforcing because a negative affective state is removed[5]. This release will be experienced as especially pleasurable and reinforcing if the change occurs rapidly, both because the contrast will be enhanced between high and low anxiety states, and because such changes will be more effective at producing greater magnitude reward-prediction errors, with concomitant phasic increases in neuromodulators such as dopamine[6,7].

RAR is a unified model for understanding all humorous phenomena, without exception. For jokes that do not appear to involve rapid releases of anxiety, these seeming exceptions are predicted to be less humorous, and thus constitute further evidence for the theory. For example, most people think that puns are only moderately funny, but puns are also only moderately effective at inducing anxiety, and most people are only moderately effective at engaging in cognitive reframing capable of rapidly releasing that anxiety.

This formulation is not intended reduce all humorous experiences to the rapid reduction of anxiety in an eliminative sense. Rather, humor is deployed in numerous and flexible ways, with each of these likely having substantial impacts on the specific ways that humor is experienced. Yet across all of these cases, the core emotion[8] may be one of rapid anxiety reduction.

## Comparisons with other theories

Several competing and complimentary theories of humor have been proposed, with RAR having the most in common with False Alarm Theory (FAT)[9], Benign Violation Theory (BVT)[10], and Cognitive Debugging Theory (CDT)[11].

### False Alarm Theory (FAT)
FAT states that humor involves a gradual build-up of expected threat, followed by a sudden non-threatening change in expectation, such that anticipation of threat is reduced. Laughter would serve as a means of informing conspecifics that they need not orient to the false alarm, and this ethological function is proposed to be the primary selective pressure leading to the origin of humor. RAR can be considered to be a more precise formulation of this model, with FAT providing an evolutionary account of selective pressures leading to the—likely brainstem mediated[12-14]—laughing reflex. However, RAR proposes humor itself may have originally evolved as a byproduct of selection for more fundamental capacities[15]. More specifically, an organism will respond with pleasure to threat reduction to the extent that associated anxiety is experienced as aversive, can be rapidly released, and thereby produces negative reinforcement. In this way, humorous experience could have evolved regardless of the social signaling value of laughter, since the underlying mechanisms are essential for basic goal-oriented behavior[16,17]. To the extent that humor is uniquely developed in humans, this may be explainable by the fact that we can so easily experience anxiety from both subtle and complex patterns, and also rapidly reduce that experience via cognitive reframing.

### Benign Violation Theory (BVT)
BVT states that humor involves the perception of a violation that is simultaneously perceived as benign. RAR can be considered to be a broader principle that explains BVT, with the addition of two crucial details:
1. The humorous effect of benign violation is mediated by anxiety reduction.
2. Humor is inversely proportional to the time it takes for the anxiety to be reduced.

To the extent that there seem to be exceptions to BVT, it is because either a) the violation fails to produce anxiety, or b) the benign reframing fails to occur rapidly.

### Cognitive Debugging Theory (CDT)
CDT states that humor is an evolved process that helps humans to maintain the data-integrity of world knowledge by rewarding them for detecting errors. Again, RAR can be considered to be a broader explanatory principle for which CDT is a special case. That is, cognitive debugging events are humorous to the extent that they involve a rapid reduction of anxiety. However, this is not the primary reason that humor evolved. As previously described, RAR considers humor to be a byproduct of organisms that a) experience anxiety as aversive, b) experience pleasure accompanying anxiety reduction, and c) are capable of manipulating their cognitive states in ways that can rapidly reduce anxiety. Rather than a specific mechanism being needed for cognitive debugging, this process of error-detection may happen automatically as part of a basic cortical algorithm that allows the brain to function as a predictive memory system[18,19]. Humorous pleasure may help to enhance debugging—and hence represent an additional selective pressure for humorous experience, particularly for organisms that depend on both precise and complex cognition[20]—but this would be just one of many dynamics that contribute to shaping humor sensitivity.

## Directions for further research

RAR appears to be the first unified theory capable of explaining the full range of humorous phenomena. The theory can be falsified if a single example is found wherein either a) substantial humor can be demonstrated without rapid reductions in anxiety, or b) rapid releases in anxiety can be demonstrated without the experience of humor. Granted, this falsifiability is limited by the facts that a) it is difficult to prove the absence of phenomena, and b) the amount of requisite rapidity may vary from person to person.

Event-related fMRI could potentially be used to identify particular neural systems associated with humorous experience, some of which may include brainstem, amygdala[21], anterior insula[22,23], and anterior cingulate cortex[24]. However, given the time-dependency of humor, better data may be obtained with the superior temporal resolution—albeit inferior spatial resolution—of electroencephalography or magnetoencephalography.

More compelling support may involve using different theories to construct jokes, and seeing which jokes are most reliably perceived as humorous. For example, emphasizing anxiety and timing should produce superior jokes as compared with BVT.

If humor is such a universal experience, and if it can be explained by such a simple formula, then why do so few people think of humor in this way? This question may be answered by the very thing that allows us to take pleasure in humor in the first place. That is, if people are negatively reinforced by the reduction of anxiety, and if contemplating anxiety makes people anxious, then people will be shaped in ways that make them less likely to contemplate anxiety-based models. Further, contemplating anxiety will make jokes less humorous if this activity either a) reduces the amount of induced anxiety, or b) reduces the degree to which that anxiety can be rapidly reduced. For these reasons, many people are likely to find RAR to be counter-intuitive, or at least until they become comfortable with the idea [25,26].